\documentclass{eas}
\usepackage{graphicx}
\begin{document}
\runningtitle{A. Sozzetti: Planetary Systems with $\mu$as Astrometry}
\title{Detection and Characterization of Planetary Systems with $\mu$as Astrometry } 
\author{A. Sozzetti}
\address{INAF - Osservatorio Astronomico di Torino, 
Strada Osservatorio, 20 - 10025 Pino Torinese Italy}
%
%
\begin{abstract}

Astrometry as a technique has so far proved of limited utility when employed as either a follow-up 
tool or to independently search for planetary mass companions orbiting nearby stars. 
However, this is bound to change during the next decade. 
In this review, I start by summarizing past and present efforts to detect planets via milli-arcsecond 
astrometry. Next, I provide an overview of the variety of technical, statistical, and astrophysical 
challenges that must be met by future ground-based and space-borne efforts in order to achieve 
the required degree of astrometric measurement precision. Then, I discuss the planet-finding capabilities 
of future astrometric observatories aiming at micro-arcsecond precision, with a particular focus on their 
ability to fully describe multiple-component systems. I conclude by putting astrometry 
in context, illustrating its potential for important contributions to planetary science, as a 
complement to other indirect and direct methods for the detection and characterization of planetary systems.

\end{abstract}
\maketitle
\section{Introduction}

The present-day catalog of extrasolar planets\footnote{See for example Jean Schneider's Extrasolar 
Planet Encyclopedia at {\tt http://exoplanet.eu}} includes a panoply of systems containing more than 
one planetary companion, up to a maximum of five in the case of the 55 Cnc system (Fischer et al. 2007). The astounding diversity 
of multiple systems encompasses some extreme configurations discovered by means of the light-time-travel 
technique around pulsars (Wolszczan \& Frail 1992), post-AGB stars (Silvotti et al. 2007), SdB+M dwarf eclipsing binaries 
(Lee et al. 2009), or by direct imaging of young stars and their dusty disks (Kalas et al. 2008; Marois et al. 2008). 
However, decade-long, high-precision (1-3 m s$^{-1}$) radial-velocity surveys have so far contributed the bulk of 
the well-characterized multiple-planet systems orbiting normal stars in the solar neighborhood known to-date. 
Their intriguing properties provide fundamental clues and insights for improved understanding of the 
complex processes of planetary systems formation, early orbital evolution, and long-term dynamical interaction. 
For example, multiple systems appear to be quite common. Recent estimates indicate that $\sim30\%$ of 
stars with planets have more than one planetary mass companion (integrating over all spectral types 
and for nearby dwarfs within 200 pc). Planetary systems exhibit great dynamical diversity, with several 
families identified, which include hierarchical systems, secularly interacting systems, and systems 
in mean motion resonances (e.g., Go\'zdziewski et al. 2008, and references therein). 
Most recently, the HD 45364 two-planet systems was shown by Correia et al. (2009) to be locked in a 3:2 
resonance, analogous to the one formed by Neptune and Pluto in our Solar System. 
Planetary systems appear to have different orbital elements distribution functions with respect to those 
of single-planet systems. In addition, distributions for systems containing only low-mass (Neptune and Super-Earths) 
planets may also differ from those of systems containing gas giants. Finally, there are hints that planet 
frequency $f_p$ may also be a different function of the host star's properties ($M_\star$, [Fe/H]) 
in single- and multiple-planet systems (for comprehensive summaries of the characteristics of multiple-planet 
systems detected by radial-velocity surveys see for example Wright et al. (2009) and the reviews by 
J. Wright and S. Udry in this volume). The observational data on multiple systems have important implications 
for the proposed models of formation and early evolution of planetary systems (for reviews see for example 
Lissauer \& Stevenson (2007), Durisen et al. (2007), Nagasawa et al. (2007), and references therein), 
provide important clues on the relative role of several proposed mechanisms of dynamical 
interactions between forming planets, gaseous/planetesimals disks, and distant companion stars 
(for reviews see for example Papaloizou et al. (2007), Levison et al. (2007), Ford \& Rasio (2008), 
and references therein), and allow to measure the likelihood of formation and survival of terrestrial planets 
in the Habitable Zone\footnote{For any given star, the region of habitability is defined 
as the range of orbital distances at which a potential water reservoir, the primary
ingredient for the development of a complex biology, would be
found in liquid form (e.g., Kasting et al. 1993)} of the parent star (Menou \& Tabachnik 2003; 
Jones et al. 2005; Hinse et al. 2008, and references therein). 

Multiple-planet systems are thus clearly excellent laboratories to search for fossil evidence of formation 
and dynamical evolution mechanisms. However, given the present limitations in our ability to elucidate in a 
unified manner the various phases of the complex processes of planet formation and evolution, some of 
the key questions on the physical characterization and architecture of planetary systems 
(how many dynamical families can be identified? Are their orbits coplanar? 
What is the origin of their eccentricities? Are the parameters' distribution functions and $f_p(M_\star$, [Fe/H]) 
actually different for single- and multiple-planet systems?) still await a clear answer. 
To this end, help from future data, obtained with a variety of techniques, over a wide range of 
observing wavelengths, both from the ground and in space, will prove invaluable. 

As for the most successful of indirect detection techniques, Doppler surveys 
are extending their time baseline and/or are achieving higher velocity precision ($\leq 1$ m s$^{-1}$, see for 
example Pepe \& Lovis 2008), to continue searching for planets at increasingly 
larger orbital distances (e.g., Wright et al. 2007) and with increasingly smaller masses 
(e.g., Mayor et al. 2008). During the next decade, RV surveys will get close to answering 
the hot question of how common are planetary systems with architectures similar to the Solar System. 
Ultimately, the limiting factor may not be the intrinsic stability of new-generation spectrographs, 
but rather the primary stars themselves, through astrophysical noise sources such as stellar surface activity, 
rotation, and acoustic p-modes. These problems are already limiting severely Doppler surveys from 
investigating the existence of giant planets orbiting stars significantly departing from our Sun in 
age, mass, and metal content. 

While ground-based wide-field photometric transit surveys are allowing us to unveil 
fundamental properties of strongly irradiated giant planets (Charbonneau et al. 2007, 
and references therein), the Kepler (Borucki et al. 2003) 
and CoRoT (Baglin et al. 2002) missions are designed to photometrically detect transiting Earth-sized 
planets in the Habitable Zone of solar-type host stars, providing the first measure of the
occurrence of rocky planets and ice-giants. The exquisite photometric 
precision and the long time baseline of the measurements will allow to look for additional companions 
in systems where one planet is found to be transiting, thanks to the possible detection of tiny 
variations in the predicted time of transit center induced by the gravitational perturbation 
of one or more outer companions, not necessarily transiting (for a review of the transit timing 
method and its potential see M. Holman, this volume). However, the host stars will reside at
typical distances beyond 250 pc, making imaging and spectroscopic follow-up of the
planets difficult. The prospects for detailed characterization of giant planets and Super Earths 
transiting nearby solar-type as well as cool stars are tied to the approval of proposed 
all-sky surveys in space (e.g., TESS), and to the possible success of upcoming 
ground-based photometric searches for transiting rocky planets aroud M dwarfs (e.g., MEarth).

Gravitational microlensing surveys from the ground within the next decade have 
the potential to deliver a complete census of the cold planet population down to $\sim10$ $M_\oplus$ 
orbiting low-mass stars at separations $a > 1.5$ AU. Proposed microlensing observatories in space (e.g., MPF) 
could extend the census to planets of $\sim 1$ $M_\oplus$ with separations exceeding 1 AU 
(e.g., Gaudi 2007, and references therein). 
We note however that observations with this technique are non-reproducible and follow-up analyses 
are virtually impossible (the detected systems typically residing at over 1 kpc from the Sun), 
thus such findings will mostly have statistical value but will help little toward the physical 
characterization of planetary systems. 

During the next twenty years, the prospects are becoming increasingly ``bright'' for 
the direct detection of exoplanets and the spectroscopic characterization of their atmospheres 
using techniques to spatially or temporally separate them from their parent stars. 
Data from upcoming and proposed observatories (for a review see for example Beuzit et al. 2007, 
and references therein) for visible-light, near- and mid-infrared imaging and spectroscopy 
and equipped with single- and multiple-aperture telescopes from the ground (e.g., VLT/SPHERE, ELT/EPICS) 
and in space (e.g., JWST, SPICA, Darwin and TPF-C/I avatars) will completely transform our view of the nature 
of planetary systems. 

In this review paper I will focus on what the contribution of astrometry from the ground
and in space will be to the astrophysics of planetary systems. I will provide an historical 
perspective on past efforts to detect planets with astrometry, I will address some of the most 
relevant challenges to be faced in the transition from milli-arcsecond (mas) to micro-arcsecond ($\mu$as) 
astrometry, and I will conclude with a discussion on future prospects, by putting this technique 
in perspective with other planet-detection methods.

\section{Blunders and Successes of mas Astrometry}

Similarly to the spectroscopic technique, astrometric measurements can detect the stellar 
wobble around the system barycenter due to the gravitational perturbation of nearby planets. 
The main observable (assuming circular orbits) is the `astrometric signature', i.e. 
the apparent semi-major axis of the stellar orbit:
\begin{equation}\label{sign}
\alpha=\left(\frac{M_p}{M_\odot}\right)\left(\frac{M_\odot}{M_\star}\right)
\left(\frac{a_p}{1\,\mathrm{AU}}\right)\left(\frac{\mathrm{pc}}{d}\right)\,\,\mathrm{arcsec}
\end{equation}
However, by reconstructing the orbital motion in the plane of the sky, 
astrometry alone can determine the entire set of seven orbital elements, 
thus breaking the $M_p\sin i$ degeneracy intrinsic to Doppler measurements and allowing to 
derive an actual mass estimate for the companion. In multiple systems, astrometric measurements 
can determine the mutual inclination angle between pairs of planetary orbits:
\begin{equation}\label{inclrel}
\cos i_\mathrm{rel} = \cos i_\mathrm{in}\cos i_\mathrm{out}+ \sin
i_\mathrm{in}\sin i_\mathrm{out} \cos(\Omega_\mathrm{out}-
\Omega_\mathrm{in}),
\end{equation}
where $i_\mathrm{in}$ and $i_\mathrm{out}$, $\Omega_\mathrm{in}$
and $\Omega_\mathrm{out}$ are the inclinations and lines of nodes
of the inner and outer planet, respectively. Thus, meaningful estimates can be obtained 
of the full three-dimensional geometry of any planetary system, without restrictions on the 
orbital alignment with respect to the line of sight. 

\subsection{Ground-Based Astrometry}

Shortly after the end of World War II, Otto Struve (1952) had already summarized the merit of 
searching for planets using precision radial-velocities, transit photometry, and astrometry. 
In his own words, ``one of the burning questions of astronomy deals with the frequency of 
planet-like bodies in the galaxy which belong to stars other than the Sun''. At that time, 
interest in the problem had been stimulated by the pre-war `discoveries' of planet-like companions 
around 61 Cygni, and 70 Ophiuchi by Strand (1943), and Reuyl \& Holmberg (1943), who had presented  
astrometric measurements based on long-term time-series of photographic plates. These announcements 
had indeed been interpreted as supporting evidence for proposed theories on the origin of the Solar 
System (Alfv\'en 1943) and speculations on the frequency of planetary systems in space (Jeans 1943). 

The evidence for planetary companions detected by ground-based astrometry around 61 Cyg and 70 Oph 
has long been proved incorrect (e.g., Heintz 1978). These two early examples are not the only ones. 
The two most famous decades-long ``quarrels'' concern Barnard's Star and Lalande 21185. The best-known 
is the long-term effort to detect planets around Barnard's star by van de Kamp (1963, 1969a, 1969b, 
1975, 1982). The claim was not confirmed by Gatewood \& Eichhorn (1973), while Hershey (1973), 
Heintz (1976), and finally Croswell (1988) showed that van de Kamp's ``detections'' were the likely 
result of instrumental errors. As for the planetary companion to Lalande 21185, the announcement 
by Lippincott (1960a, 1960b) and Hershey \& Lippincott (1982) was initially not confirmed by 
Gatewood (1974) and Gatewood et al. (1992), while in the most recent contribution to the subject 
Gatewood (1996) claimed instead one or even two planets could possibly be orbiting the star. In both 
these cases, the jury is still out.  

\subsection{Hipparcos Astrometry}

Prompted by the success of Doppler surveys for giant planets of
nearby stars and by the need to find a method to break the $M_p-i$ degeneracy
intrinsic to radial-velocity measurements, several authors have re-analyzed in recent years the Hipparcos
Intermediate Astrometric Data (IAD), in order to either detect
the planet-induced stellar astrometric motion of the bright hosts, most of which
had been observed by the satellite, or place upper limits to the
magnitude of the perturbation, in the case of no detections. The Hipparcos 
IAD have been re-processed alone, or in combination with either the
spectroscopic information or with additional ground-based
astrometric measurements.

The Hipparcos IAD have been used to put upper limits on the size of the 
astrometric perturbations by Perryman et al. (1996) and by Zucker \& Mazeh
(2001), who could rule out at the $\sim 2$-$\sigma$ level the hypothesis of 
low-mass stellar companions disguised as planets for over two dozen objects.
Preliminary astrometric masses for $\sim30$ Doppler-detected planets were announced 
about a decade ago by several authors (Mazeh et al. 1999; Zucker \& Mazeh 2000; 
Gatewood et al. 2001; Han et al. 2001). On the one hand, the suprising conclusion of these works is that 
a significant fraction ($\sim 40\%$) of the planet candidates are instead stars,
and the remainder sub-stellar companions are in most cases brown dwarfs rather
than planets. The results stem from the derivation of a vast majority
of quasi-face-on orbits, i.e. with $i$ on the order of only a few degrees. 
On the other hand, Pourbaix (2001), Pourbaix \& Arenou (2001), and later Zucker \& Mazeh (2001) 
demonstrated that the statistical significance of most of the Hipparcos astrometric orbits is questionable 
at best, and that the systematically very small inclination angles can arise as
an artifact of the fitting procedures utilized to dig out signals below the noise level of the 
Hipparcos data. Interestingly enough, the only system with an Hipparcos orbit deemed 
statistically acceptable by Pourbaix \& Arenou (2001) which corresponded to an M-dwarf companion, $\varrho$ CrB (Gatewood et al. 2001), 
was then called into question by recent high-resolution infrared spectroscopic measurements (Bender et al. 2005). 
Successful Hipparcos astrometric orbital solutions for a few Doppler-detected systems containing 
companions with minimum masses close to and slight above the dividing line between planets and brown dwarfs have 
recently obtained by Reffert \& Quirrenbach (2006) and Sozzetti \& Desidera (2009).

\subsection{HST/FGS Astrometry} 

A firm (3-5 $\sigma$ level) upper mass limit of $\sim30$ $M_J$ 
on the mass of a spectroscopically detected extrasolar planet was 
placed by McGrath et al. (2002), who failed to reveal astrometric motion
of the $M_p\sin i = 0.88$ $M_J$ object on a 14.65-day orbit in the
$\varrho^1$ Cnc multiple-planet system using narrow-field relative HST/FGS astrometry. 
Finally, the first undisputed value of the actual mass of a Doppler-detected planet 
was obtained by Benedict et al. (2002) who derived the perturbation size, inclination angle, and
mass of the outer companion in the multiple-planet system GJ 876 
from a combined fit to HST/FGS astrometry and high-precision radial-velocities. 
In recent years, HST/FGS astrometry has helped determining the actual mass of 
the Neptune-mass planet in the $\varrho^1$ Cnc system, under the assumption of 
coplanarity (McArthur et al. 2004), has allowed to measure the mass of the long-period 
planet in orbit around $\varepsilon$ Eri (Benedict et al. 2006), and has permitted to 
reveal the nature of the companion to HD 33636 as an M dwarf rather than a massive planet (Bean et al. 2007).

Until the advent of $\mu$as-level precision astrometric facilities from the ground and in space, 
mas astrometry with HST/FGS will continue to deliver important results. Data have already 
been gathered and are being analyzed in order to determine the actual masses of the Doppler-detected 
planets HD 47536b, HD 136118b, HD 168443c, HD 145675b, and HD 38529c (Benedict et al. 2008). 
Furthermore, HST/FGS observations are also being collected for a handful of multiple systems, including 
HD 128311, HD 2020206, $\mu$ Ara, $\gamma$ Cep and $\upsilon$ And, with the aim of 
measuring directly the degree of coplanarity among detectable components. Most recently, 
the first coplanarity test has been carried out by Bean \& Seifahrt (2009) in the case of the two 
outer planets in the GJ 876 systems using a combination of Doppler measurements and HST/FGS 
astrometry, albeit with the crucial help of \apriori\, dynamical considerations.

\section{The Challenges of $\mu$as Astrometry}

The state-of-the-art astrometric precision is nowadays set to $\sim 1$ mas 
by Hipparcos and HST/FGS (see \S~\ref{spaceobs}). By looking at Eq.~\ref{sign}, 
one realizes how the magnitude of the perturbation induced by 
a 1 Jupiter-mass planet in orbit at 5 AU around a 1-M$_\odot$ star at 10 pc from the Sun 
is $\alpha\simeq 500$ $\mu$as. For the same distance and primary mass, 
a `hot Jupiter' with $a_p=0.01$ AU induces $\alpha=1$ $\mu$as, and an Earth-like planet ($a_p=1$ AU) 
causes a perturbation $\alpha=0.33$ $\mu$as. 
One then understands why astrometric measurements with mas precision have so far
proved of limited utility when employed as either a follow-up tool 
or to independently search for planetary mass companions orbiting 
nearby stars. Indeed, $\mu$as astrometry is almost coming of age, provided the demanding technological 
and calibration requirements to achieve the required level of measurement precision are met. 

Sozzetti (2005) has provided a review of methods and instrumentation to detect and characterize 
planetary systems with astrometry. I summarize here the main points, focusing in particular 
on the challenges inherent to correctly modeling astrometric measurements in which 
multiple planetary signals are present. 

\subsection{Correcting for Astrophysical Effects}

With the goal of achieving $\mu$as-level precision, astrometric observations may have to be corrected 
first for a variety of effects that modify the apparent position of the target. 
These can be classical in nature 
or intrinsically relativistic, and can be due to a) the motion of the observer (e.g., aberration), 
b) secular variations in the target space motion with respect to the observer (e.g., perspective acceleration), 
or c) the gravitational fields of massive bodies in the vicinity of the observer (light deflection). 
Taking into account such effects is particularly important for global astrometric 
measurements such as those that will be carried out by Gaia (less so for differential measurements). 
Indeed, several attempts have been made in the past years (Brumberg 1991; 
Klioner \& Kopeikin 1992; de Felice et al. 1998, 2001, 2004, 2006; de Felice \& Preti 2006, 2008; 
Vecchiato et al. 2003; Klioner 2003, 2004; Crosta et al. 2003) to develop schemes for the reduction 
of astrometric observations at the $\mu$as precision level directly within the framework of
General Relativity, either employing non-perturbative approaches or the PPN formalism (Will 1993). 
A model of relativistic astrometry based on the Klioner (2003) PPN formulation is 
considered the baseline for the reduction of Gaia data by the Gaia Data Processing and Analysis Consortium. 

\subsection{Noise Sources}

The astrometric measurement process is carried out in the presence of observational error sources (both random and 
systematic) which depend on the mode of operation (wide-, narrow-angle, or global astrometry), 
operational wavelength (visible or near-infrared) 
and the instrument (monolithic or diluted configuration) used to carry out the measurements.  

As for instrumental errors, for single-dish architectures the most technologically challenging 
to deal with will be the ability to achieve location accuracies of $\sim1/1000$ of a pixel for CCD detectors, 
and the capability to minimize geometric distortions of optical systems. 
The diffraction-limited image quality afforded by adaptive optics systems 
modifies the relative importance of these error terms to some degree (Cameron et al. 2009).
For interferometers, both accuracies of tens of picometers on the position of the delay lines 
and positional stabilities of of $\sim 10$ nm on internal optical pathlengths, 
in order to ensure maintenance of the fringe visibility, will have to be achieved (see Sozzetti 2005, 
and references therein). For both architectures, non-uniform system throughput due to the time evolution of 
the optical parameters (Gai \& Cancelliere 2008) could also induce significant degradation 
in the achievable astrometric precision. 

For ground-based instrumentation, the atmosphere constitutes an
additional source of noise through both its turbulent layers (a random component) and due  
to the differential chromatic refraction (DCR) effect (a systematic component). The limitations due 
to the former effect can be overcome to a significant degree utilizing diluted rather than monolithic 
configurations (Shao \& Colavita 1992). The latter has been proven by several studies to be often 
the predominant limitation to precision astrometry from the ground (e.g., Monet al. 1992; Lazorenko 2006). 

For space-borne observatories, additional random and systematic error sources must be taken into
account, which are introduced by the satellite operations and environment. The most relevant 
class of uncertainties can be broadly defined in terms of attitude errors 
(induced by solar wind, micrometeorites, particle radiation, radiation pressure, thermal drifts and 
spacecraft jitter), 
for which {\it ad hoc} calibration procedures must be implemented case by case. In \S~\ref{spaceobs} 
I describe in some detail how these problems have been approached in the past, and how they are 
being addressed for future programs.

Finally, for $\mu$as-level astrometric precision several `astrophysical' noise 
sources (due to the environment or intrinsic to the target) begin to play a significant role. 
These include the dynamical effect of previously unknown stellar companions, astrometric `jitter' 
induced by stellar surface activity (spots, flares), and by the presence of a circumstellar protoplanetary disk 
(the motion of the disk center of mass provoked by the excitation of spiral density waves 
by an embedded planet, dynamical perturbations due to the disk self-gravity if it's marginally unstable, 
and time-variable, asymmetric starlight scattering). Such effects have been studied in detail by 
several authors (Sozzetti 2005, and references therein; Eriksson \& Lindegren 2007). 
The general picture is that, unlike the radial-velocity case, 
$\mu$as-level astrometry is significantly less affected by the above astrophysical noise sources. 

\subsection{Modeling Planetary Systems} 

There are many difficulties inherent to the problem of astrometric orbit fitting for planetary systems. 
Orbital fits require highly non-linear fitting procedures, with a large number of 
model parameters: For a system with $n_p$ planetary companions, the model parameters of the orbital 
fit will be $5+7\times n_p$, i.e. the five classic astrometric parameters + seven orbital elements per 
object, not including additional solutions for the space motion of reference stars. 
Particularly for $n_p>1$ , several complex problems must be overcome in order to 
successfully fit multiple Keplerian orbits. These include, for example, a) the trade-off between accuracy 
in the determination of the mutual inclination angles between pairs of planetary orbits, 
single-measurement precision and redundancy in the number of observations with respect 
to the number of estimated model parameters (typically $n_\mathrm{obs} >> n_p$), 
b) the merging of radial velocity + astrometric datasets in combined solutions to improve 
the quality of orbit reconstruction and mass determination, c) the careful assessment of 
the relative robustness and reliability of different procedures for multiple-planet orbital fits, 
and d) the challenge to correctly identify signals with amplitude close to the measurement uncertainties 
(those typically produced by terrestrial planets), particularly in the presence of larger 
signals induced by other companions and/or sources of astrophysical noise (due to, 
e.g., stellar surface structure or variably illuminated disks) of comparable magnitude. 
All the above issues can have a significant impact in any attempt to gauge 
the actual capabilities of an astrometric instrument aiming at detecting and characterizing 
planetary systems, and terrestrial planets in particular. 

Recent studies (Casertano et al. 2008; Traub et al., this volume; Wright et al., in 
preparation; Sozzetti et al., in preparation) have begun investigating in detail some of 
the abovementioned critical aspects. First of all, in these works several independent 
algorithms for single- and multiple-component orbital fits have been implemented and utilized. 
These robust, global least-squares fitting procedures adopt, for example, partial linearizations of 
the multi-body Kepler problem, different minimization techniques to optimally search the 
orbital parameter space, such as Bayesian inference and Markov-Chain Monte Carlo 
analysis (e.g., Ford \& Gregory 2007), or frequency decomposition (e.g., Konacki  et al. 2002), 
and are used to carry out single- and multiple-planet orbital solutions on simulated 
astrometric data alone as well as on combined high-precision astrometric+RV datasets. 
For example, within the context of a double-blind tests campaign for planet deteciton with 
SIM-Lite (see \S~\ref{SIM} and Traub et al. in this volume), the figure of merit 
utilized in the (iterative) minimization process is the sum of the separate $\chi^2$ values:
\begin{equation}
\chi^2_\mathrm{tot}=
\sum\left(\frac{x^M-x^\star}{\sigma_\mathrm{x}}\right)^2+
\sum\left(\frac{y^M-y^\star}{\sigma_\mathrm{y}}\right)^2+
\sum\left(\frac{RV^M-RV^\star}{\sigma_\mathrm{RV}}\right)^2,
\end{equation}
where $x^\star$, $y^\star$ are the position differences 
(in rectangular coordinates) between a target and a reference object as 
measured by SIM-Lite and $RV^\star$ the radial velocity of the primary, respectively, $\sigma_\mathrm{x}$, 
$\sigma_\mathrm{y}$, and $\sigma_\mathrm{RV}$ are the associated uncertainties, and 
$x^M$, $y^M$, and $RV^M$ are the predicted values based on the orbital elements 
at each iteration, and the sum is carried out over all observations of each kind. 
A particularly valuable aspect of this approach is the ability to carry out combined 
(three-dimensional) solutions even when one type of observation (astrometry or spectroscopy, 
and sometimes both) has insufficient coverage 
for an independent fit. Simultaneous orbital fits can particularly strenghten 
the determination of orbital elements and masses of the companions because 
they fully exploit the redundancy constraints available from both types of data. 

Second, in the above studies a number of statistical indicators (periodogram FAPs, F-tests, MLR tests) 
have been used, and the criteria for regulating relative agreement among them established, for a most 
thorough, robust assessment of the quality and reliability of the 
orbital solutions obtained. In addition, detailed understanding of the statistical 
properties of the uncertainties associated with the model parameters have been obtained, 
through improved understanding of the relative merit of error analyses based on e.g., 
covariance matrices, $\chi^2$ surface mapping, and bootstrapping procedures. 
This approach is necessary because of the 
large parameter space to be investigated and due to the fact that for highly non-linear 
fitting procedures the statistical properties of the solutions are not at all trivial 
(and significantly differ from those of linear models). 
Doppler surveys are already facing the challenges of fitting multiple-component 
orbits, and it is not uncommon to find strong disagreement between solutions 
(and sometimes number of planets detected!) presented by different teams 
(e.g., Butler et al. 2006; Gregory 2007). 

Finally, for multiple-component orbital fits, future work will also focus on 
the inclusion of N-body integrators, in order to account for possibly significant dynamical interactions. 
In multi-planet systems, planet-planet interactions can significantly alter the RV and/or 
astrometric signature of the system. In such cases where interactions are important 
(as for the GJ 876 planetary system), 
a full dynamical (Newtonian) fit involving an n-body code must be used to properly model 
the data and to ensure the short- and long-term stability of the solution.

\subsection{Achieving $\mu$as Astrometry: Ground-Based Experiments}

The possibility of achieving $\mu$as-level astrometric precision from the ground 
with monopupil telescopes at visible wavelengths 
has been tested by numerous past experiments (e.g., Gatewood 1987; Han 1989; Monet et al. 1992), 
essentially confirming the theoretical limits imposed by atmospheric noise (e.g., Lindegren 1980) that 
hamper the ability to significantly push below uncertainties on the order of 
$\sim 1$ mas, particularly in the long term. The best short-term precision achieved with the 5-m 
Palomar telescope (Pravdo \& Shaklan 1996) motivated the Stellar Planet Survey (STEPS), an 
astrometric survey for very low-mass companions to nearby M-dwarf stars. The long-term noise floor is 
$\sim1$ mas (Pravdo et al. 2005). 
Recently theoretical studies which include adaptive optics observations and symmetrization of the reference frame 
to remove low-frequency components of the image motion spectrum and improve image centroid 
at the data-reduction stage predict improved performances for 10-m class telescopes 
(Lazorenko \& Lazorenko 2004). 
The theoretical predictions have recently begun to be put under experimental tests. 
Lazorenko (2006), Lazorenko et al. (2007), R\"oll et al. (2008) and Cameron et al. (2009) have 
employed relatively narrow-field imagers on the Palomar and VLT telescopes to demonstrate short-term 
$100-300$ $\mu$as precision (see also the papers by R\"oll et al. and Helminiak in this volume). 
Pioneering studies of coronagraphic astrometry (Digby et al. 2006), which encompass 
the use of an occulting mask in conjunction with adaptive optics devices, have investigated several methods 
(centroiding, instrument feedback, analysis of point-spread function symmetry) to carry out 
precision astrometry of an occulted star. Preliminary results show performances not below mas-level precision. 

The promise of long-baseline optical/infrared interferometry for
high-precision astrometry (Shao \& Colavita 1992) has been tested by a number of experiments
in the past (e.g., Colavita et al. 1994) The best performances have been achieved by the Palomar Testbed Interferometer 
in phase-referencing mode, with $\sim100$ $\mu$as short-term accuracy for $\sim30^{\prime\prime}$ 
binaries (Lane et al. 2000) and $20-50$ $\mu$as for sub-arcsec binaries (Lane \& Muterspaugh 2004) within the 
context of the Palomar High-precision Astrometric Search for Exoplanet Systems (PHASES) program. 
The PHASES observations have been able to exclude tertiary companions with masses as small as 
a few Jupiter masses with $a< 2$ AU in several binary systems (Muterspaugh et al. 2006). 

\begin{figure}
\centering
\includegraphics[width=.75\textwidth,angle=0.]{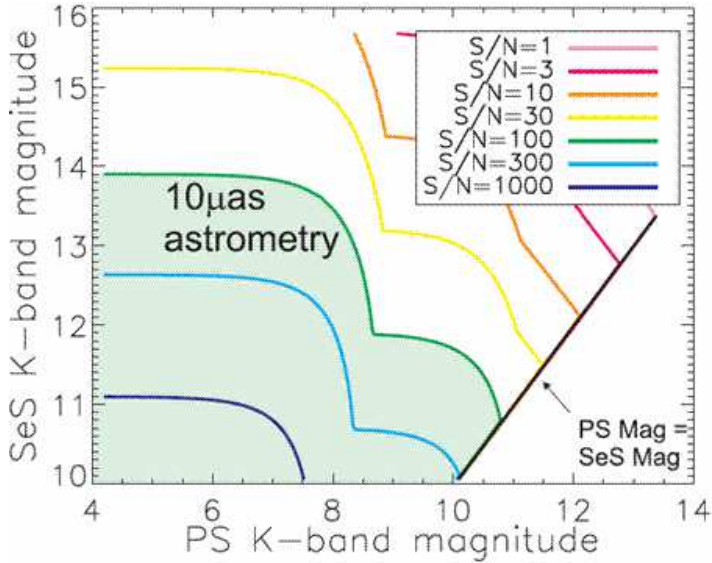}
\caption{Dependence of astrometric accuracy of PRIMA with two ATs on stellar brightness (K-band magnitude) for
$10^{\prime\prime}$ separation between primary (PS) and secondary (SeS) star and 30 min integration time. 
Note that this figure is the result of an error budget simulation, based on PRIMA 
subsystem specifications, but without knowing the actual throughput 
and performance of all components. The actual sensitivity of PRIMA will 
be verified on real stars during the commissioning of the instrument in  2009. 
{\it Credits: Launhardt et al. 2008}.}
\label{PRIMA}
\end{figure}

The predicted astrometric performances of large aperture, ground-based interferometers equipped with adaptive
optics systems, such as Keck-I (Ragland et al. 2008) and the VLTI (Glindemann et al. 2000), 
hold promise to approach the actual limits of this technique from the ground. 
The ASTrometric and phase-Referenced Astronomy (ASTRA) project upgrade (Pott et al. 2008) 
of Keck-I will add dual-star capability for high sensitivity observations and dual-star astrometry. 
The new facility, to be ready within 2 years' time, has quoted limiting magnitudes 10 mag and 
15 mag in $K$-band for narrow-angle astrometry at the $\sim100$ $\mu$as level 
between pairs of objects separated by $< 20-30$ arcsec. 
PRIMA, the instrument for Phase-Referenced Imaging and Micro-arcsecond Astrometry at the VLTI 
(Delplancke 2008), is currently being integrated and tested at Paranal, with 
science operations presently scheduled for early 2010. Similarly to Keck-I, 
the PRIMA/VLTI facility is designed to perform narrow-angle interferometric astrometry with two ATs 
of a target and one reference star separated by up to 1 arcmin. Launhardt et al. (2008) have 
shown that, using error budget simulations based on PRIMA subsystem specifications, 
in order to achieve the astrometric precision of 10 $\mu$as in 30 min 
integration time, the primary target and the reference (of limiting $K$-band brightness 
8 mag and 14 mag, respectively) should not be separated by more than $\sim20^{\prime\prime}$ (see Figure~\ref{PRIMA}). 

The Exoplanet Search with PRIMA (ESPRI) Consortium (Launhardt et al. 2008) 
will carry out a three-fold observing program focused on the astrometric 
characterization of known radial velocity planets within $\leq 200$ pc from the Sun, 
the astrometric detection of low-mass planets around nearby stars of any spectral type within $\approx15$ 
pc from the Sun, and the search for massive planets orbiting young stars with ages in the 
range $5-300$ Myr within $\sim100$ pc from the Sun. The initial target list includes 
$\sim900$ stars, but the number of good references will likely reduce the batch by 
almost an order of magnitude, resulting in a tentative target list of $\sim100$ stars 
to be observed with PRIMA during the 5-yr duration of the survey 
(Launhardt et al. 2009, in preparation). An extensive program of preparatory observations to 
characterize suitable references and weed out from the target list excessively 
active stars and short-period binaries is underway, which includes high dynamic 
range near-infrared photometric imaging and spectroscopic observations. 

\subsection{Achieving $\mu$as Astrometry: Space-Borne Observatories}\label{spaceobs}

Relative, narrow-angle astrometry from space has been performed so
far with the Fine Guidance Sensors aboard HST, while global
astrometric measurements have been carried out for the first time by Hipparcos.

For HST/FGS astrometry with respect to a set of reference objects near the target 
(within the $5\times 5$ arcsec instantaneous field of view of FGS), 
pre-launch error budget estimates (e.g., Bahcall \& O'Dell 1980) predicted 1-2 mas 
single-measurement precision down to $m_v\approx 16$. This performance level has been 
demonstrated by e.g., Benedict et al. (1994, 1999) who, in the data reduction of 
the two-dimensional interferometric measurements, devised {\it ad hoc} calibration
and data reduction procedures to remove a variety of random and systematic error
sources from the astrometric reference frame (intra-observation spacecraft jitter, 
temperature variations and temperature-induced changes in the secondary mirror position, 
constant and time-dependent optical field angle distortions, intra-orbit drifts, 
lateral color corrections). The limiting factor is the spacecraft jitter. 
A single-measurement precision below 0.5-1 mas is out of reach for HST/FGS.

For the Hipparcos all-sky survey, the achievable precision (Lindegren 1989) in the along-scan direction 
had received pre-launch estimates of $\sim 1$ mas on bright objects, with degradation 
of a factor of several at the magnitude limit of the survey, depending on location on 
the sky (given the pre-determined scanning law adopted for the satellite). 
Actual sky-averaged uncertainties confirmed the predictions, with typical uncertainties 
degrading from $\sim1.0$ mas for $m_v < 7$ to $\sim 4.5$ mas for $m_v\geq 11$ 
(e.g., Kovalevsky 2002), achieved with a calibration and iterative reduction scheme 
in several steps (Lindegren \& Kovalevsky 1989) that allowed to successfully 
derive values of positions, proper motions, and parallaxes simultaneously
for $\sim 120,000$ stars by bridging one-dimensional angular
measurements along the satellite's instantaneous scanning direction
into a global astrometric solution over the whole celestial sphere. 
Without the presence of the atmosphere, and similarly to HST/FGS, the best-achievable 
single-measurement precision is limited by the uncertainties in the determination
of the along-scan attitude.

The ability to suppress systematics by at least two orders of
magnitude for a space-borne instrument is a major technological
goal. Both SIM-Lite and Gaia promise to achieve this level of
astrometric precision. The newly redesigned SIM-Lite has a shorter baseline (6 m) with respect to the 
old SIM configuration (10 m), and it replaces a guide 
interferometer with a telescope star tracker. It will deliver better 
than 1 $\mu$as narrow-angle astrometry in 1.5 hr integration time (Goullioud et al. 2008) 
on bright targets ($m_v \leq 7$) and moderately fainter references ($m_v\simeq 9-10$). 
For this purpose, an accuracy on the position of the delay lines of a few tens of pm 
with a 6-m baseline must be achieved (Zhai et al. 2008). Furthermore, a positional
stability of internal optical pathlengths of $\sim 10$ nm is
required, in order to ensure maintenance of the fringe visibility
(Goullioud et al. 2008). For Gaia, the success in meeting the goal of 
$\approx 10$ $\mu$as single-measurement astrometric precision to hunt for planets around bright stars
($m_v < 13$) will depend on a) the ability to attain CCD
centroiding errors not greater than 1/1000 of a pixel in the
along-scan direction (Gai et al. 2001) and b) the capability to
limit instrumental uncertainties (thermo-mechanical stability of
telescope and focal plane assembly, metrology errors in the
monitoring of the basic angle) and calibration errors
(chromaticity, charge transfer inefficiency, 
satellite attitude, focal plane-to-field coordinates transformation) 
down to the few $\mu$as level (e.g., Perryman et al. 2001).

\section{The Potential of $\mu$as Astrometry}

A number of authors have tackled the problem of
evaluating the sensitivity of the astrometric technique required to 
detect extrasolar planets and reliably measure their orbital elements and 
masses (Sozzetti 2005, and references therein). Those works mostly relied on simplying assumptions 
with regard to a) the error models to be applied to the data (e.g., 
simple gaussian distributions, perfect knowledge of the instruments) and 
b) the analysis procedures to be adopted for orbit reconstruction (mostly ignoring the problem 
of identifying adequate configurations of starting values from scratch). 
The two most recent exercises on this subject (Casertano et al. 2008; Traub et al., this volume) 
have revisited earlier findings using a more realistic double-blind protocol. 
In this particular case, several teams of ``solvers'' were handled simulated datasets of 
stars with and without planets and independently defined detection tests, 
with levels of statistical significance of their own choice, and orbital fitting algorithms, 
using any local, global, or hybrid solution method that they devised was best. The solvers 
were provided no \apriori\, information on the actual presence of planets around a given target. 

\subsection{Gaia DBT} 

\begin{figure}
\centering
\includegraphics[width=.24\textwidth,angle=270.]{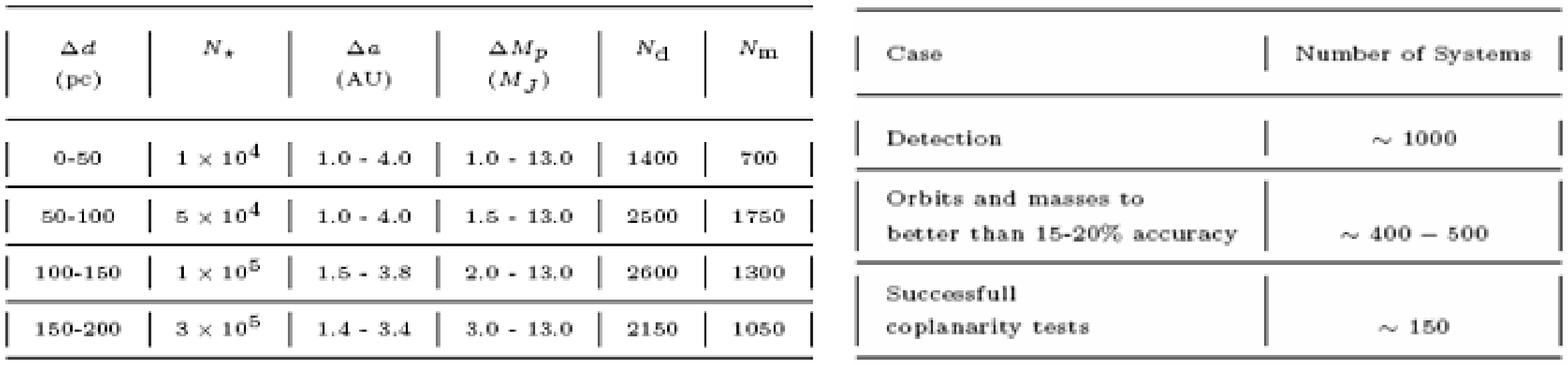}
\caption{Left: Number of giant planets that could be detected and measured by Gaia, as a 
function of increasing distance. Starcounts are obtained using the Besancon model of 
stellar population synthesis (Bienaym\'e et al. 1987), while the Tabachnik \& Tremaine (2002) model 
for estimating planet frequency as a function of mass and orbital period is utilized. 
Right: Number of multiple-planet systems that Gaia could potentially detect, measure, 
and for which coplanarity tests could be carried out successfully. {\it Credits: Casertano et al. 2008}.}
\label{Gaia}
\end{figure}

In the large-scale, double-blind test (DBT) campaign carried out to estimate the potential of 
Gaia for detecting and measuring planetary systems, Casertano et al. (2008) showed that $a)$ 
planets with $\alpha\simeq 6\sigma$ (where $\sigma$ is the single-measurement error) 
and orbital periods shorter than the nominal 5-yr mission 
lifetime could be accurately modeled, and $b)$ for favorable configurations of two-planet systems 
with well separated periods (both planets with $P\leq 4$ yr and $\alpha/\sigma\geq 10$, 
redundancy over a factor of 2 in the number of observations) it would be possible to carry out 
meaningful coplanarity tests, with typical uncertainties on the mutual inclination angle of $\leq 10$ deg. 
Both subtle differences as well as significant discrepancies were found in the orbital solutions carried 
out by different solvers. This constitutes further evidence that the convergence of non-linear 
fitting procedures and the quality of orbital solutions (particularly for multiple systems and for 
systems with small astrometric signals) can both be significantly affected by the choice of the 
starting guesses for the parameters in the orbital fits, by the adoption of different statistical 
indicators of the quality of a solution, and varied levels of significance of the latter.

Overall, the authors concluded that Gaia could discover and measure massive giant planets ($M_p \geq 2-3$ $M_\mathrm{J}$)
with $1<a<4$ AU orbiting solar-type stars as far as the nearest star-forming regions,
as well as explore the domain of Saturn-mass planets with similar
orbital semi-major axes around late-type stars within 30-40 pc. These results can be turned
into a number of planets of given mass and orbital separation that can be detected and measured by Gaia, 
using Galaxy models and the current knowledge of exoplanet frequencies. By inspection of the tables in 
Figure~\ref{Gaia}, one then finds that Gaia's main strength will be its ability to measure accurately 
orbits and masses for thousands of giant planets, and to perform coplanarity measurements for a few 
hundred multiple systems with favorable configurations. 

\subsection{SIM-Lite DBT}\label{SIM}

\begin{figure}
\centering
\includegraphics[width=.32\textwidth,angle=270.]{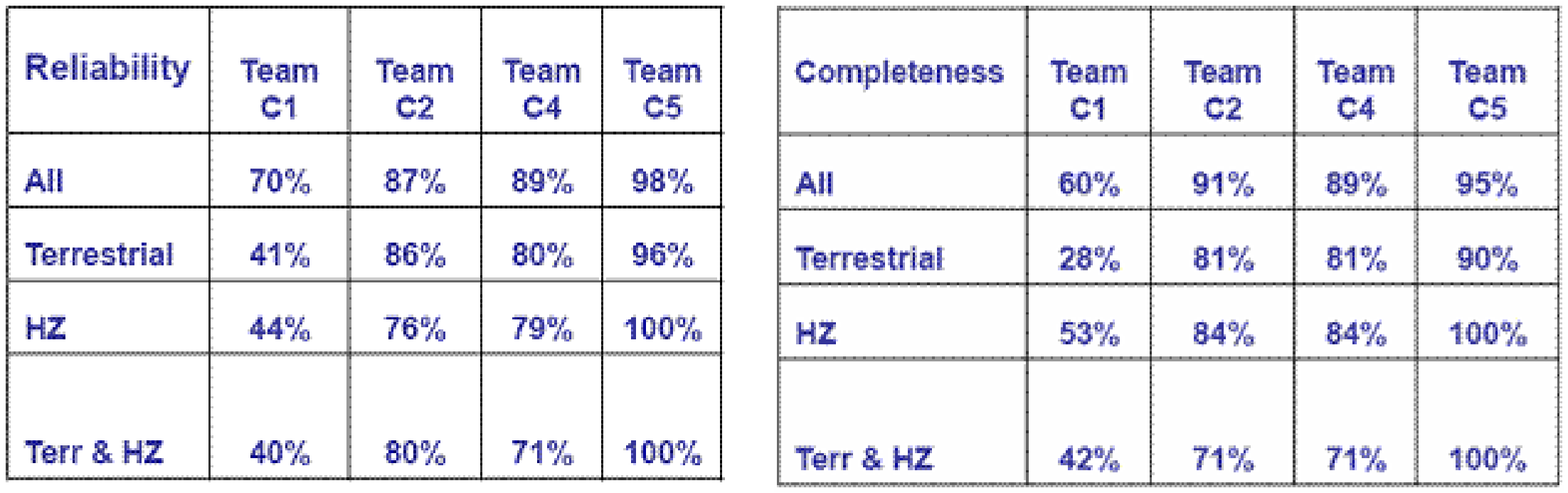}
\caption{Left: reliability of detection (number of detected planets divided by the number 
of detected planets + the number of false alarms) vs. planet type, for various teams contributing 
to the SIM-Lite double-blind tests campaign. Right: completeness (number of detected planets divided 
by number of detectable planets) vs. planet type, for the same teams. {\it Credits: Wes Traub.}}\label{SIMLite}
\end{figure}

The second study carried out in double-blind mode set out to answer a number of questions relating to 
the detectability of Earth-like planets (terrestrial masses and Habitable-Zone orbits) in multi-planet systems, 
using a combination of SIM-Lite astrometry and ground-based radial-velocity observations. 
As discussed in detail by Traub et al. (this volume), the theoretical predictions of required 
signal-to-noise ratio to detect and measure planetary systems were proven correct (similarly to the 
Casertano et al. (2008) analysis). Both reliability of detection (i.e., the probability that a 
detection based on a periodogram analysis is true and not a false alarm) and completeness 
(i.e., the probability that a planet will be detected) were gauged, and found in very good agreement 
with the expectations. Similarly to Casertano et al. (2008), this study also highlighted how different 
detection and orbit fitting algorithms, given the same datasets, can perform in measurably different ways 
(see Figure~\ref{SIMLite}).  

Overall, it was demonstrated that other planets do not significantly interfere with the detection of 
terrestrial planets in the Habitable Zone, and to reach the sensitivity needed to detect Earth-like planets 
$\sim40\%$ of a 5-yr SIM-Lite mission with $\sigma=0.82$ $\mu$as is required, 
with the additional help of 15-yr baseline RV data. 

A second phase of the SIM-Lite double-blind excercise is underway, in which the study will be 
extended to real stars in the SIM-Lite target list, all detections will be subjected to additional 
statistical tests and long-term stability analyses in order to more robustly assess the reliability 
of any detection and the confidence on multi-component orbital solutions, and requirements will be 
placed on the accuracy on masses and orbital parameters necessary to usefully inform direct imaging surveys 
about the epoch and location of maximum brightness, in order to estimate optimal visibility. 
The latter issue is of particular relevance in light of concerns recently raised by Brown (2009) about 
the ultimate requirements on astrometry to support the planning of direct observations
of terrestrial habitable planets, indicating $\mu$as-level precision may not suffice.  

\section{Summary}

An improvement of 2-3 orders of magnitude in achievable measurement precision, down to the 
$\mu$as level, would allow this technique to achieve in perspective the same
successes of the Doppler method, for which the improvement from the km s$^{-1}$
to the m s$^{-1}$ precision opened the doors for ground-breaking results in 
exoplanetary science. Indeed, $\mu$as astrometry is almost 
coming of age. Provided the demanding technological and calibration requirements to achieve the 
required level of measurement precision are met, future observatories at visible and near-infrared wavelengths, 
using both monolithic as well as diluted architectures from the ground (VLTI/PRIMA, Keck-I) 
and in space (Gaia, SIM-Lite) hold promise for crucial contributions to many aspects of planetary 
systems astrophysics (formation theories, dynamical evolution, internal structure, detection 
of Earth-like planets), in combination with data collected with other indirect and direct techniques.

\begin{figure}
\centering
\includegraphics[width=.75\textwidth]{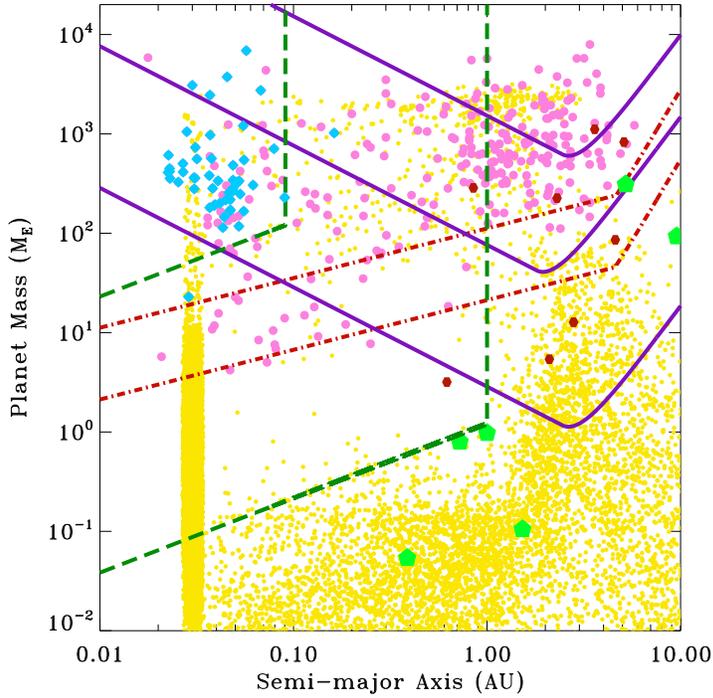}
\caption{Exoplanets discovery space for the astrometric, Doppler, and transit techniques. 
Detectability curves are defined on the basis of a 3-$\sigma$ criterion for signal detection. 
The upper and middle blue curves are for Gaia astrometry with $\sigma_\mathrm{A} = 10$ $\mu$as, 
assuming a 1-$M_\odot$ G dwarf primary at 200 pc and a 0.4-$M_\odot$ M dwarf at 25 pc, respectively, 
while the lower blue curve is for SIM-Lite astrometry of a 1-$M_\odot$ star at 10 pc with 
$\sigma_\mathrm{A} = 0.8$ $\mu$as. For both Gaia and SIM-Lite, survey duration is set to 5 yr. The radial velocity 
curves (red lines) assume $\sigma_\mathrm{RV} = 3$ m s$^{-1}$ (upper curve) 
and $\sigma_\mathrm{RV} = 1$ m s$^{-1}$ (lower curve), $M_\star = 1 M_\odot$, and 10-yr survey duration. 
For visible-light transit photometry (green curves), the assumption are $\sigma_V = 5\times10^{-3}$ mag (upper curve) and 
$\sigma_V = 1\times10^{-5}$ mag (lower curve), $S/N = 9$, $M_\star=1$ $M_\odot$, $R_\star = 1$ $R_\odot$, 
uniform and dense ($>> 1000$ datapoints) sampling. 
Pink dots indicate the inventory of Doppler-detected exoplanets as of December 2008. Transiting systems 
are shown as light-blue filled diamonds, while the red hexagons are planets detected by microlensing. 
Solar System planets are also shown as green pentagons. The yellow small dots represent a theoretical 
distribution of masses and final orbital semi-major axes from Ida \& Lin (2008).\label{space}}
\end{figure}

Figure~\ref{space} shows the $M_p$-$a$ diagram with
the plotted present-day and achievable sensitivities of transit photometry and
radial-velocity, and with the expected SIM-Lite and Gaia detection thresholds
at 10 pc, 25 pc, and 200 pc, respectively. The presently known planets detected by the various 
methods are also shown, along with the predicted distribution of recent models (Ida \& Lin 2008). 
At first glance, one could get the impression that the impact of astrometric measurements 
(except for those obtained by SIM-Lite around the nearest stars) may not bear great potential. 
However, the relative importance of different planet detection techniques should not 
be gauged by looking at their discovery potential {\it per se}, but rather in connection 
to outstanding questions to be addressed and answered in the science of planetary systems. 

Some of the most important issues for which $\mu$as astrometry will play a key role 
in the next decade include: a) a significant refinement of our understanding of 
the statistical properties of extrasolar planets: for example, the predicted Gaia 
database of several thousand extrasolar planets with well-measured properties will allow 
to test the fine structure of giant planet parameters
distributions and frequencies, and to investigate their possible changes
as a function of stellar mass, metal content, and age with unprecedented resolution; 
b) crucial tests of theoretical models of gas giant planet formation and migration: 
for example, specific predictions on formation time-scales and the role of varying 
metal content in the protoplanetary disk will be probed with unprecedented statistics thanks
to the thousands of metal-poor stars and hundreds of young stars
screened for giant planets out to a few AUs by Gaia, VLTI/PRIMA, and SIM-Lite; 
c) key improvements in our comprehension of importants aspects of the 
astrophysics of multiple-planet systems: for example, coplanarity tests for hundreds 
of multiple-planet systems will be carried out with Gaia, SIM-Lite, and VLTI/PRIMA, 
and this, in combination with data available from Doppler measurements and transit timing, 
could allow to discriminate between various proposed mechanisms for eccentricity
excitation, thus significantly improving our comprehension of the role of dynamical interactions 
in the early as well as long-term evolution of planetary systems; 
d) an important contribution to the understanding of direct detections of giant extrasolar planets: 
for example, accurate knowledge of all orbital
parameters and actual mass are essential for understanding the thermophysical conditions 
on a planet and for determining its visibility. 
Actual mass estimates and full orbital geometry determination for suitable
systems (with typical separations $> 0.1^{\prime\prime}$), obtained by means of 
high-precision astrometric measurements (with Gaia, SIM-Lite, and VLTI/PRIMA), 
will inform direct imaging surveys about the epoch and location of maximum brightness, in order 
to estimate optimal visibility, and will help in the modeling and interpretation of the 
phase functions and light-curves of giant planets (the first prediction about 
where and when to look for the planet $\epsilon$ Eridani b was recently made by Benedict et al. (2006) 
using HST/FGS astrometry); e) the collection of essential supplementary data for the optimization 
of the target lists of future observatories (e.g., Beichman et al. 2007) aiming at the direct 
detection and spectroscopic characterization of terrestrial, 
habitable planets in the vicinity of the Sun. For example, astrometry of all nearby stars within 25 pc of the 
Sun with 1-10 $\mu$as precision (with SIM-Lite and Gaia in space, and VLTI/PRIMA from the ground) 
will provide a comprehensive database of F-G-K-M stars screened for Jupiter-, Saturn-, and Neptune-mass 
companions out to several AUs. These observations would help probing the long-term dynamical stability 
of their habitable zones, where terrestrial planets may have formed, and maybe found, 
complementing on-going efforts of Doppler surveys and studies of exo-zodiacal cloud emission 
(with ground-based facilities such as Keck-I, VLTI, and LBTI). 

\section{Acknowledgements}

I am especially grateful to the Conference organizers for giving me the opportunity to write this review. 
I am indebted to Mario Lattanzi, Stefano Casertano, and Alessandro Spagna for guiding my first steps 
in the realm of high-precision astrometry. 
I warmly thank Fritz Benedict, Wes Traub, Jason Wright, and Ralf Launhardt for very useful discussions and 
for providing material ahead of publication. Financial support from the Italian Space Agency through ASI 
contract I/037/08/0 (Gaia Mission - The Italian Participation to DPAC) is gratefully acknowledged. 


\end{document}